\documentclass[twocolumn,prb,showpacs,preprintnumbers,amsmath,amssymb]{revtex4-1}
\usepackage{graphicx}
\usepackage{epstopdf}
\usepackage{dcolumn,color}
\usepackage{bm}
\let\vec=\boldsymbol

\begin{document}

\title{Microscopic analysis of the magnetic form factor in low-dimensional cuprates}

\author{V. V. Mazurenko$^{1}$,  I. V. Solovyev$^{1,2}$ and A. A. Tsirlin$^{3,4}$}
\affiliation{$^{1}$Theoretical Physics and Applied Mathematics Department, Ural Federal University, 620002 Ekaterinburg, Russia \\
$^{2}$ Computational Materials Science Unit, National Institute for Materials Science, 1-1 Namiki, Tsukuba, Ibaraki 305-0044, Japan\\
$^{3}$National Institute of Chemical Physics and Biophysics, 12618 Tallinn, Estonia\\
$^{4}$Experimentalphysik VI, Universit\"at Augsburg, 83615 Augsburg, Germany}

\begin{abstract}
We analyze the magnetic form factor of Cu$^{2+}$ in low-dimensional quantum magnets by taking the metal-ligand hybridization into account explicitly. In this analysis we use the form of magnetic Wannier orbitals, derived from the first-principles calculations, and identify the contributions of different atomic sites. Having performed local density approximation calculations for cuprates with different types of ligand atoms, we discuss the influence of the on-site Coulomb correlations on the structure of the magnetic orbital. The typical composition of Wannier functions for copper oxides, chlorides and bromides is defined and related to features of the magnetic form factor. We propose easy-to-use approximations of the partial orbital contributions to the magnetic form factor in order to give a microscopic explanation for the results obtained in previous first-principles studies.
\end{abstract}

\maketitle

\noindent {\it Introduction.} The hybridization between localized $3d$ states of a transition metal and $p$-states of ligand atoms plays a crucial role in electronic and magnetic properties of transition metal oxides. For instance, the famous concept of charge transfer compounds \cite{Zaanen} is based on the fact that the insulating gap is defined by the energy required to excite an electron from the bonding ligand-like to antibonding metal-like states.  In contrast to Mott insulators, where the lowest-energy excitation is between the metal-like states, such an excitation occurring within the same electronic configuration does not involve any change in Coulomb energy, $U$.

The mixing of metal and ligand states is also of great importance for magnetic properties of transition metal oxides.  First, the hybridization is responsible for delocalization of the magnetic moment that is described by a wave function containing, e.g., copper and ligand states. This effect is most pronounced in copper oxides, chlorides and bromides in which the half-filled $3d$ orbital of $x^2-y^2$ symmetry lies higher in energy, points toward ligand atoms and overlaps with $p$-orbitals of ligands. Thus the magnetic moment observed in experiments is associated with a cluster of atoms, which should be taken into account when constructing a realistic magnetic model of a particular material. 
Additionally, the metal-ligand hybridization triggers long-range superexchange interactions due to the electron hopping. The strength of the magnetic coupling in this case is controlled by the angle of the metal-ligand-...-ligand-...ligand-metal bond and the nature of the ligand atom. \cite{PRB}

Numerical methods based on the density functional theory are a powerful tool to study the covalent effects in low-dimensional magnets and to amend common rules often used for the microscopic analysis of magnetic couplings, such as Goodenough-Kanamori-Anderson rules. \cite{Goodenough}
For instance, the ab-initio results reported in Ref.\onlinecite{Rosner} have shown that the large size of Cl and Br atoms leads to a strong overlap of ligand $p$ - orbitals along magnetic pathways and amplifies next-nearest-neighbor couplings in spin-chain compounds. The increased Cu-ligand hybridization eventually results in a non-Goodenough-Kanamori regime when the ferromagnetic coupling is enhanced as the Cu-L-Cu angle increases. The authors of the work have also revisited the critical metal-ligand-metal angles separating the antiferromagnetic and ferromagnetic regimes. 

Experimentally, the information concerning the magnetic couplings in a low-dimensional spin system can be extracted from the inelastic neutron scattering (INS) measurements that directly probe the spin-spin correlation functions. To reproduce the experiment, one can use the text-book expression containing the spin-spin correlation function (see for instance Ref.\onlinecite{White}). Normally, the spin density in this approach is assumed to be fully localized, and the magnetic form factor is associated with that for the Cu$^{2+}$ ion. \cite{Garlea, Cavadini} However, several cases in the literature \cite{Crowe,Kernavanois} indicate the importance of the ligand contributions that are crucial to reproduce the experimental INS spectra. 

Important results concerning the effect of covalent bonding on the magnetism of copper oxides were obtained in Ref.\onlinecite{Walters}. It was shown that the larger extent of the magnetic Wannier function in the real space, as triggered by the copper-oxygen hybridization, leads to the shrinking and additional redistribution of the form factor in the wave-vector
space.  The account of such a redistribution is important for an accurate description of the magnetic intensities in INS spectra.

In this paper, we perform a microscopic analysis of partial orbital contributions to the magnetic form factor by taking into account the hybridization between metal and ligand states. We show that the theoretical spectrum can be divided into metal-metal, metal-ligand and ligand-ligand contributions that are directly related to the structure of the Wannier function describing the spatial distribution of the magnetic moment.  The typical compositions of the Wannier orbital in the case of copper oxide, chloride and bromide are determined from first-principles calculations.   
Then we propose an approximation of the magnetic form factor that takes into account the Cu$^{2+}$ and ligand contributions.  Test simulations for Li$_2$CuO$_2$ and TlCuCl$_3$ are performed. \\

\noindent {\it The structure of the Wannier function}. Theoretical description of the metal-ligand hybridization in modern materials  is a complicated problem that can be solved on the different levels. First, one can use the molecular ligand theory where the metal and ligand states are taken into account on equal footing in the model electronic Hamiltonian. However, such a formulation complicates the consideration since the model contains a lot of parameters that have to be defined. \cite{yildrim} For instance, there are metal-ligand hopping integrals, energies of metal $3d$ and ligand $p$-states and different terms of the Coulomb interaction matrix. The construction of an effective spin Hamiltonian requires a fourth-order perturbation theory and its formulation becomes rather complicated for the metal-ligand-ligand-metal bonds. 

A more elegant approach to describe the magnetic interactions in $3d$ metal oxides was formulated by Anderson.\cite{Anderson} It is based on Wannier functions that provide a very compact and accurate local representation of the electronic structure. This approach gives direct insight into the nature of chemical bonding and excitation states of materials.  

\begin{figure}[!h]
\includegraphics[angle=0,width=0.25\textwidth]{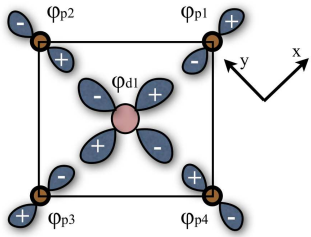}
\caption{(Color online) Schematic representation of a typical local geometry in low-dimensional cuprates. The copper atom with the active orbital of  $x^2-y^2$ symmetry is positioned at the center of the square formed by ligand atoms. The signs + and - denote the phases of the atomic wavefunctions.}
\end{figure}

In this paper, we use a Wannier-function-based approach in order to describe the magnetic properties of  copper-based low-dimensional magnets with a typical local geometry presented in Fig.1. In this case, the typical antibonding Wannier function describing the magnetic moment can be written in an atomic basis as
\begin{eqnarray}
W (\vec r) = \alpha \phi_{d} (\vec r) + \beta \sum_{p=1}^{4}  \phi_{p} (\vec r) + O(\vec r),
\end{eqnarray}
where $\phi_{d}$ and $\phi_{p}$ are atomic copper and ligand wave functions with the phases defined in Fig.1. $\alpha$ and $\beta$ are the contributions of the copper and ligand states to the Wannier function, and $O(\vec r)$ describes the contributions from the rest of the crystal.  The ligand contribution to the Wannier function is related to the metal-ligand hopping, $t_{dp}$ and can be estimated as $\beta= \frac{t_{dp}}{\Delta}$, where $\Delta$ is the charge-transfer gap (the energy splitting between the metal and ligand states).\cite{Solovyev} As we will show below, in the case of the  low-dimensional cuprates about 90 \% of the electron density is concentrated at the cluster containing the metal and four ligand atoms.  

The Wannier function can be calculated  by a projection procedure \cite{proj} that is based on the local density approximation \cite{OKA} (LDA). In this approach, the Bloch function is projected onto the set of the localized atomic-like orbitals.  Since in $3d$ compounds
the electron or spin density has strong spatial dependence, the LDA breaks down in
the description of ground state properties of transition
metal oxides. In practice, it means that one should include explicitly the on-site Coulomb interactions for localized $3d$
states, for instance by using the local spin density approximation plus Hubbard $U$ approach (LSDA+$U$). \cite{Anisimov}  

The $\alpha$ and $\beta$ coefficients in Eq.(1) for several representative types of cuprates were calculated by using linear-muffin-tin-orbital atomic sphere approximation (TB-LMTO-ASA) method within LDA and LSDA+$U$ approximations.\cite{OKA, Anisimov} In these calculations,  we used known crystal structure data, Ref.\onlinecite{struct_Li2CuO2, struct_SrCuBO, struct_BaCuSiO, struct_TlCuCl3, struct_BPCB, struct_telluride} for Li$_2$CuO$_2$, SrCu$_{2}$(BO$_3$)$_2$, BaCuSi$_{2}$O$_{6}$, TlCuCl$_{3}$, (C$_{5}$H$_{12}$N)$_2$CuBr$_4$ and Cu$_2$Te$_2$O$_5$X$_{2}$(X=Cl,Br), respectively. The results are summarized in Table I. One can see that the localization of the Wannier function strongly depends on the type of the ligand atom. In the case of low-dimensional copper oxides, the $\beta$ coefficient varies from 0.2 to 0.3. \cite{Mazurenko1, Mazurenko2} The copper compounds with chlorine and bromine ligands demonstrate much stronger metal-ligand hybridization, which can lead to $\alpha$ and  $\beta$ that are close to 0.6 and 0.4, respectively. 

To extract the values of $\alpha$ and $\beta$ from LSDA+$U$ calculations, we used the obtained magnetic moments of the metal and ligand atoms in the ferromagnetic configuration.  The choice of the ferromagnetic state with maximal total magnetization is important, since it reveals the magnetic moment of the ligand atoms that can be hidden in an antiferromagnetic configuration. In these calculations the values of the on-site Coulomb interaction $U_d$ =10 eV and intra-atomic exchange interaction $J_H$ = 1 eV were used for simulating the correlation effects in the $3d$ band of copper. These parameters were calculated by using constrained-LDA procedure.\cite{Korotin} $\alpha^2$ and $\beta^2$ are associated with the magnetic moments of metal and ligand atoms, respectively.  

\begin{table}[!h]
\centering
\caption{Contributions of metal and ligand atomic orbitals to magnetic Wannier function in cuprates with different types of ligand atoms. $\alpha_{LDA}$ ($\beta_{LDA}$) and $\alpha_{U}$ ($\beta_{U}$) correspond to the results of LDA and LSDA+$U$ calculations, respectively. The values in the parentheses for the mixed-type cuprates correspond to the contribution of chlorine or bromine atoms.}
\label{betaJ}
\begin {tabular}{cccccc}
\hline
type & compound  & $\alpha_{LDA}$ &$\beta_{LDA}$ & $\alpha_{U}$ &$\beta_{U}$ \\
  \hline
  oxide & Li$_2$CuO$_2$  & 0.73 & 0.29 &0.82  &  0.27 \\
  oxide & SrCu$_{2}$(BO$_3$)$_2$ & 0.65 & 0.27 & 0.84 & 0.22 \\
  oxide & BaCuSi$_{2}$O$_{6}$     &  0.65   & 0.27   & 0.8 &  0.26 \\
  chloride & TlCuCl$_{3}$                 &  0.64  & 0.36      & 0.63 &  0.33  \\
  bromide & (C$_{5}$H$_{12}$N)$_2$CuBr$_4$       & 0.59    & 0.34 & 0.6 &   0.37     \\
  mixed & Cu$_2$Te$_2$O$_5$Br$_2$ & 0.69 & 0.25 (0.41) & 0.82 & 0.23  (0.31) \\
  mixed & Cu$_2$Te$_2$O$_5$Cl$_2$ & 0.71 & 0.25 (0.36) &0.85  & 0.22 (0.26)\\
  \hline
\end {tabular}
\end {table}

Having compared LDA and LSDA+$U$ compositions of the Wannier function, we conclude that the main changes concern the metal contribution. 
The consideration of the Coulomb correlations in oxides leads to an increase of metal contribution by 12-23 \%, which is related to the additional downward shift of occupied $3d$ states in LSDA+$U$. 

The information concerning the structure of the Wannier function plays an important role in the analysis of magnetic and electronic properties of strongly correlated materials. Our previous results have shown that the exchange interaction in case of nearly 90$^{\circ}$ metal-ligand-metal bond has a ferromagnetic contribution that is driven by the intra-atomic Hund's coupling on the ligand atom. \cite{Mazurenko1,Mazurenko2, Rosner} The resulting exchange interaction is very sensitive to the shape of the Wannier function. As we will show below, the properties of the magnetic form factor and, therefore, the INS spectra also depend on the spread of the magnetic orbital. Thus the definition of the Wannier function plays a crucial role for modeling magnetic properties of low-dimensional magnets.  \\

\noindent {\it Magnetic form factor.} The magnetic scattering of unpolarized neutrons by the spin magnetic moment of the electrons  is proportional to the spin-spin correlation function,\cite{Kaplan}
\begin{eqnarray}
\mathcal{S} (\vec q, \omega) = \int d \vec r d \vec r' dt \langle \hat {\vec S}(\vec r, t) \hat {\vec S}(\vec r')  \rangle e^{-i\vec q (\vec r -\vec r') + i \omega t},
\end{eqnarray}
where $\hat {\vec S} (\vec r)$ is the spin density operator, and  the momentum transfer $\vec q$ is the difference between incident and scattered neutron wave vectors, $\vec q = \vec k -\vec k'$. Without external magnetic field and anisotropy $\hat S^{z} (\vec r) \hat S^{z} (\vec r')$, $ \hat S^{+} (\vec r) \hat S^{-} (\vec r')$ and $\hat S^{-} (\vec r) \hat S^{+}(\vec r')$ contributions to the cross-section are equal to each other. Therefore, we can consider only one of them.  In the Wannier function basis, the $z$-component of the spin density operator can be expressed in the following form:
\begin{eqnarray}
\hat S^z (\vec r, t) = \frac{1}{2} \sum_{i  \sigma} W^*_{i}(\vec r) W_{i} (\vec r) \sigma \hat d^+_{i \sigma} (t) \hat d_{i \sigma} (t),
\end{eqnarray}
where the spin index $\sigma=\pm 1$, $W_{i} (\vec r)$ is the Wannier function centered at the site $i$, $d^{+}_{i \sigma}$ is the creation operator. We assume that the system can be simulated with an one-band model in the spirit of our Wannier analysis.
Thus the general formula Eq.(2) can be rewritten in the following form
\begin{eqnarray}
\mathcal{S} (\vec q, \omega) = \int_{-\infty}^{\infty} dt  \sum_{ij}  \langle \hat S^z_i (t) \hat S^z_j \rangle  e^{-\omega t}  \rho_{i} (\vec q) \rho_{j} (-\vec q), 
\end{eqnarray}
where $\hat S^z_i $ is the spin operator of the $i$th site, and the $\vec q$-dependent electron density is given by
\begin{eqnarray}
\rho_i (\vec q) = \int d \vec r e^{-i \vec q \vec r} W^*_{i} (\vec r) W_{i} (\vec r), 
\end{eqnarray} 
which is essentially the magnetic form factor.
It can be directly calculated by using the Fourier transform of the electron density obtained from the first-principles calculations. The examples of the magnetic form factors for Li$_2$CuO$_2$ and TlCuCl$_3$ are given in Fig.2. To demonstrate the effect of covalent bonding, we compare the obtained spectra with the magnetic form factor of the Cu$^{2+}$ ion that was calculated by using the 3-Gaussian approximation proposed in Ref.\onlinecite{PJBrown}. There is a strong redistribution of the oxide and chloride spectra, which fully agrees with the results of work Ref.\onlinecite{Walters}. One also observes an oscillating behavior  of the calculated form factors of the copper systems.

\begin{figure}[!h]
\includegraphics[angle=0,width=0.45\textwidth]{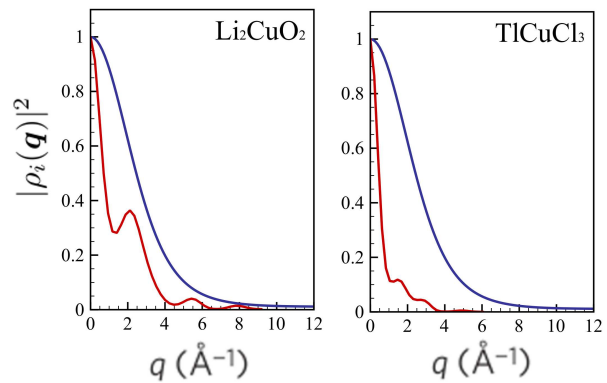}
\caption{(Color online) Magnetic form factors (red lines) calculated for cuprates with different types of the ligand atoms. The momentum transfer $\vec q$ is chosen to be along the copper-ligand bond. The blue line corresponds to the magnetic form factor of Cu$^{2+}$ ion.}
\end{figure}

To provide a microscopic explanation of the obtained first-principles results, we decompose the Wannier function with the copper and ligand contributions, Eq.(1). The density can be written in the following form
\begin{eqnarray}
\rho_i (\vec q) = \alpha^2 \int d \vec r \rho^{d}_{i} (\vec r) e^{-i \vec q \vec r} + \beta^2 \sum_{p} \int d \vec r \rho_{i}^{p} (\vec r) e^{-i \vec q \vec r},
\label{rho} 
\end{eqnarray} 
where $\rho^{d}_{i} (\vec r) = \phi^{*}_{d} (\vec r - \vec R_{i}) \phi_{d} (\vec r - \vec R_{i})$, $\rho^{p}_{i} (\vec r) = \phi^{*}_{p} (\vec r - \vec R_{i}-\vec R_{p}) \phi_{p} (\vec r - \vec R_{i} - \vec R_{p})$ and $\vec R_{p}$ is the position of the ligand atom with respect to the $i$th metal atom. Here $\phi_{d}$ and $\phi_{p}$ are atomic-like wave functions of metal and ligand atoms, respectively.  The resulting form factor has the metal-metal ($\rho^{d}_{i} (\vec q) \rho^{d}_{i} (-\vec q) $), metal-ligand ($\rho^{d}_{i} (\vec q) \rho^{p}_{i} (-\vec q) $) and ligand-ligand ($\rho^{p}_{i} (\vec q) \rho^{p}_{i} (-\vec q) $) contributions. The metal-metal contribution is associated with the magnetic form factor for the Cu$^{2+}$ ion. \cite{Freltoft, Crowe}  

Since there is a power dependence in Eq.(6) on the $\alpha$ and $\beta$ coefficients, the resulting spectrum is very sensitive to the structure of the magnetic Wannier functions. 
In the remaining part of the paper, we propose a simple approximation to the ligand contributions. On one hand, it gives us an opportunity to perform a microscopic analysis of the calculated magnetic form factors (Fig.2).  On the other hand, our approach can be used for a fast analysis of the experimental data avoiding first-principles calculations.  Below, we consider different approximations for the ligand contribution to the form factor and analyze the obtained partial spectra. 

\begin{figure}[!h]
\includegraphics[angle=0,width=0.45\textwidth]{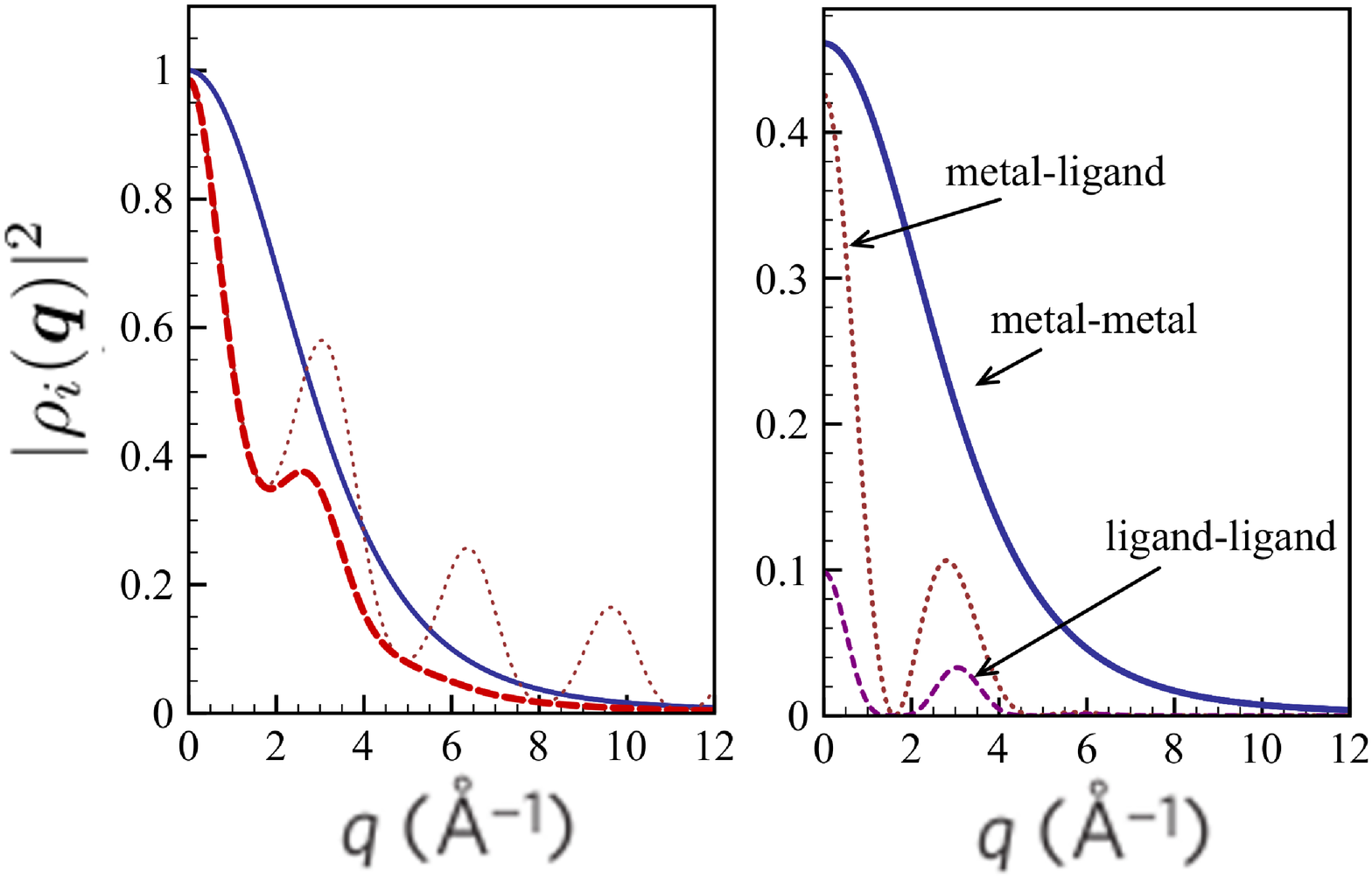}
\caption{(Color online) (Left) Comparison of the magnetic form factors of the Cu$^{2+}$ ion (solid line) and that calculated by using the delta function (dotted line) and Gaussian (dashed line) approximations for the ligand part of the electron density. (Right) Partial contributions to the magnetic form factor calculated by using the Gaussian distribution for the electron density of the ligand.}
\end{figure}

The first approximation we consider is to replace the ligand electronic density with the delta function   
 $\rho^{p}_{i} (\vec r) = \delta (\vec r -\vec R_{i} - \vec R_{p})$. A similar approximation was used in Ref.\onlinecite{Solovyev} for simulating the electron density within the Wannier function basis in the analysis of the electronic polarization in manganites. Despite of the crudeness of this approximation, it demonstrates the main effect of the oxygen states, which is the redistribution of the spectrum at small $\vec q$. Thus we obtain
\begin{eqnarray}
\rho_i (\vec q) = \alpha^2 \rho_{i}^d (\vec q) + \beta^{2} \sum_{p} e^{-i \vec q (\vec R_{i} + \vec R_{p})}
\end{eqnarray}
where $\rho_{i}^d (\vec q)$ is the form factor for the Cu$^{2+}$ ion. To test this approximation we performed the calculations of the form factor in the case of Sr$_2$CuO$_3$ described in Ref.\onlinecite{Walters}. The coefficients $\alpha$ and $\beta$ were chosen to be 0.82 and 0.28, respectively. The metal-ligand bond length is 1.95 \AA. The comparison of the results for Cu$^{2+}$ ion and our delta-function approximation is presented in Fig.3 (left, dotted line). In accordance with the results of Ref.\onlinecite{Walters}, we observe a redistribution of $|\rho(\vec q)|^2$ at small $q$. There are also strong oscillations of the spectrum at $|\vec q|>$  2 \AA$^{-1}$. They are controlled by the geometry of the metal-ligand cluster, the metal-oxygen distance, $\vec R_{p}$. 

To improve the delta-function approximation of the ligand contribution to the electron density, we follow an approach proposed in Ref.\onlinecite{Lynn}. The authors used the Gaussian distribution to describe the ligand electron density, 
\begin{eqnarray}
\rho^{p}_{i} (\vec r) = \frac{1}{\sigma \sqrt{2 \pi}} e^{\frac{-(\vec r -\vec R_{i} - \vec R_{p})^2} {2 \sigma^2}},
\end{eqnarray}
where $\sigma$ is the standard deviation.  In this case, the Fourier transform of $\rho^{p}_{i} (\vec r)$ has also a Gaussian form
\begin{eqnarray}
\rho^{p}_{i} (\vec q) = e^{- i \vec q (\vec R_{i}+ \vec R_{p})} e^{-\frac{\vec q^2 \sigma^2}{2}}.
\end{eqnarray}
Using such a form for $\rho^{p}_{i} (\vec q)$, we obtained the spectrum (Fig.3, left, dashed line) which agrees with that presented in Ref.\onlinecite{Walters}.
The analysis of the partial contributions to $|\rho_{i} (\vec q)|^2$ revealed that the electron density at small $q$ is due to the metal-ligand hybridization.
 
The last step of our investigation was to reproduce the calculated form factors for Li$_2$CuO$_2$ and TlCuCl$_3$ by using the proposed approximation. To this end,  the parameter $\sigma$ was varied in order to obtain the best fitting of the calculated spectra (Fig.4). With coefficients $\alpha$ and $\beta$ presented in Table I we obtained $\sigma_{Li_2CuO_2 }$ = 0.39 \AA \, and $\sigma_{TlCuCl_3 }$ = 3.5 \AA. It demonstrates a quantitative difference between copper oxides and chlorides on the level of the magnetic form factors and the much broader distribution of the scattering spin density in chlorides compared to oxides.

\begin{figure}[t]
\includegraphics[angle=0,width=0.45\textwidth]{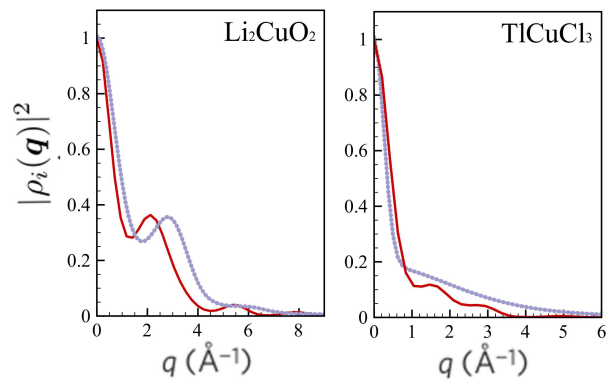}
\caption{(Color online) Fitting of the calculated form factors (Fig.2) by using the Gaussian approximation for ligand contribution, Eq.(9) proposed in this work.}
\end{figure}

The proposed approximations can be used for the analysis of non-local contributions to the spin-spin correlation function,  $\rho_{i} (\vec q) \rho_{j} (-\vec q)$.
The correction for the ligand-ligand contribution,  $\rho^p_{i} (\vec q) \rho^p_{j} (-\vec q)$ will modify the position of the INS maximum in the momentum space. The corresponding correction is proportional to the ratio $\frac{|\vec R_{p}|}{|\vec R_{ij}|}$, where $|\vec R_{ij}|$ is the metal-metal distance. We left the analysis of the non-local contributions to the magnetic form factor for a future investigation.

To conclude, we propose a simple approximation scheme for different contributions to the magnetic form factor in low-dimensional cuprates.  The main features of the spectrum, such as the redistribution and peak at $|\vec q| > 0$, are due to the metal-ligand hybridization. The structure of the Wannier function describing the magnetic moment in a particular system is related to the different contributions to the features of the magnetic form factor spectrum. The obtained results can be used not only for a microscopic explanation of the experimental INS spectra, but also for the analysis of the local geometry of copper-ligand cluster in low-dimensional magnets.  \\

\noindent {\it Acknowledgements.}
We acknowledge fruitful communication with Fr\'ed\'eric Mila, Frank Lechermann and Vladimir Dmitrienko.
The work of VVM and IVS is supported by the grant program of the Russian Science Foundation 14-12-00306. AAT acknowledges financial support of PUT733 (Estonian Research Council) and the Sofja Kovalevskaya Award of Alexander von Humboldt Foundation.

\end{document}